\documentstyle[12pt,aasms4]{article}
\begin{document}

\title{The nature of the activity in Hickson compact groups of galaxies.}

\author{Roger Coziol\altaffilmark{1}, Andr\'e L. B. Ribeiro\altaffilmark{2}}
\author{Reinaldo R. de Carvalho\altaffilmark{3}}
\and
\author{Hugo V.Capelato\altaffilmark{2}}

\altaffiltext{1}{Laborat\'orio Nacional de Astrof\'{\i}sica - LNA/CNPq, Rua Estados Unidos, 154, Bairros das Na\c{c}\~oes - 37500-000 - Itajub\'a, MG, Brasil}

\altaffiltext{2}{Divis\~ao de Astrof\'{\i}sica - INPE/MCT, C.P. 515 - 12201-970 - 
S. Jos\'e dos Campos, SP, Brasil}

\altaffiltext{3}{Observat\'orio Nacional, Rua Gal. Jos\'e Cristino, 77 - 20921-400 - Rio de Janeiro, RJ, Brasil}

\begin{abstract}

We present the results of the spectral classification of
the 82 brightest galaxies in a sample of 17 compact groups.
We verify that the AGNs are preferentially located in the most
early--type and luminous galaxies of the groups, as is usually
observed in the field.  But these AGNs also appear to be
systematically concentrated towards the central parts of the groups.
Our observations suggest a correlation between
activity types, morphologies and densities of galaxies in the compact
groups.  This is consistent with a scenario in which galaxies of compact
groups evolve by interacting with their environment and 
are currently in a quiet phase of their activity.

\end{abstract}

\keywords{galaxies: Compact groups -- 
galaxies: Seyfert --  galaxies: LINERs -- galaxies: interactions}
 
\section{Introduction}  

One of the important aspects of the study of the Hickson compact 
groups of galaxies (HCGs; Hickson 1982) resides in the attractive possibility of assessing 
the effects of strong interactions on the morphology and
stellar content of galaxies (Hickson et al. 1992).  
The available data on the HCGs however seems to present many contradictions with a fast
merging evolution scenario, suggesting that we still do not
fully understand the nature of these systems.  To shed new light on
this problem a new spectroscopic survey of faint galaxies in the
regions of Hickson compact groups was recently undertaken by de
Carvalho et al. (1997; see also Ribeiro et al. 1997) to determine the
kinematical structure of the groups. These studies show that HCGs
exhibit a variety of dynamical configurations as opposed to the
previous view that they are all isolated and high density
structures.

Spectroscopic observations of galaxies around HCGs not only allow to
establish how isolated these structures are from other structures but
also yield valuable information on the nature of the activity of the 
individual groups. Previous studies on the activity of the galaxies
in compact groups have led to contradictory conclusions.
Recently, for example, Sulentic \& Raba\c{c}a (1993) and Vegnugopal (1995)
contest the claim by Hickson et al. (1989a) that the far infrared emission is
enhanced in compact groups. Radio
observations (Menon, 1992,1995), optical spectroscopy and
imaging (Rubin et al. 1991) all suggest that tidal interactions
and mergers between compact group galaxies did happen in the past. 
However, many galaxies in the groups seem to be normal and it
is not clear which phenomenon, either starburst or AGN, is the main
source of activity observed in the groups.  One extreme example of activities
encountered in a compact group is HCG 16, which includes one Seyfert 2
galaxy, two luminous LINER galaxies and 3 starburst galaxies (Ribeiro
et al 1996). But in a sample of 17 groups, HCG 16 is the only one of
its kind, which suggests that in general the activity in the groups is
not prominent.

In this contribution we present the results of our classification of the
activity types of a sample of luminous galaxies from 17 HCGs.  We show
that a significant fraction of these galaxies display AGN activity. 
Moreover, 50\% of the AGN population in the groups as studied here are
``low--luminosity'' AGNs (LLAGNs) -- that is, faint AGNs, either Seyfert 2
or LINER, which are hidden behind the strong stellar continuum of their
host galaxy.  We also show that the AGN population always appears
segregated towards the compact cores (which generally encompass most
of Hickson's original compact groups, Ribeiro et al. 1997) whereas
starburst galaxies tend to be distributed in their external parts.

The organization of the paper is as follows. In Section 2 we define 
our sample of compact group galaxies and explain the criteria that
we used to do our classification. We also present the characteristics of the
LLAGNs in our sample and show the results of template
subtraction, which is an essential part of the 
classification scheme.  Section 2.1 presents the results
of our classification of all the emission line galaxies following two
diagnostic diagrams of line ratios. A brief discussion on the
low--luminosity nature of the LLAGNs follows in Section 2.2.  In
Section 3, we discuss the spatial and morphological distribution of the
active galaxies in the compact groups. We conclude, in Section 4, with a
brief digression on the significance of our findings for the
understanding of the nature of compact groups.

\section {Spectral classification of the AGNs candidates}

The spectra presented in this paper are part of a sample of 316
galaxies in the regions of 17 HCGs which were observed by de Carvalho
et al. (1997).  The spectra were taken at the 4m CTIO telescope using
the ARGUS fiber--fed spectrograph. The details of the instrumental
setup and data reduction are discussed by de Carvalho et al. (1997).
In this sample, 82 galaxies (of which only 67 are kinematically
assigned to the groups: see de Carvalho et al. 1997) have spectra with
sufficiently high S/N to allow a proper classification of the activity type.
Of these galaxies, 28 (34\%) present only absorption lines and 54 (66\%) present
both emission and absorption lines. In this article the non-emission
line galaxies are considered non--active. A complete analysis of the 
absorption features in these galaxies will be published
elsewhere (Coziol et al. in preparation).
This article is dedicated to the classification of the
emission--line galaxies. 
 
Luminous emission--line galaxies are usually easy to identify and
classify.  These galaxies are either starburst or AGN galaxies.  For
many galaxies in our sample, however, the usual classification criteria
cannot be applied directly, since they do not show any emission lines
(i.e. within the limits of our observations) with the exception of the
[\ion{N}{2}]$\lambda$6584 line. Figure 1 shows the spectra for these
galaxies.  The great similarity of the spectra suggests 
that either all these galaxies are of similar morphological type or
they are dominated by the same stellar populations.
Trace of ionized gaz in early--type
spirals and ellipticals was already observed 
before and it is generally suggested that a mild 
Seyfert or LINER, that is a low--luminosity AGN, could reside in the 
nuclei of all these galaxies (see Filippenko \& Sargent 1985, Phillips et
al. 1986; hereafter PJDSB).
Given the similarity of the phenomenon observed here,
we will provisionally call the subsample of the
galaxies displayed in figure 1 the LLAGNs candidates. 
The basic characteristics of the LLAGN candidates are presented in
Table 1.  The numbers in columns 1 and 2 follow the numbering used in
de Carvalho et al. (1997) and the letters in column 3 correspond to the
galaxies in the Hickson (1982) list. Redshifts determined by de Carvalho 
et al. (1997) are listed in column 4.  The B luminosities, as taken from 
Hickson et al. (1989), are listed in column 5. The absolute
magnitudes, presented in column 6, were determined assuming 
H$_0 = 75$ km s$^{-1}$ Mpc$^{-1}$.
In column 7 we give also the morphologies of the galaxies, as reported
by Mendes de Oliveira \& Hickson (1994).
 
In the LLAGN candidates, the presence of an intermediate-age
stellar population introduces strong Balmer absorption lines that
interfere with the observation of faint ionized regions. The problem of
detecting and measuring weak emission lines atop a strong stellar
continuum has already been tackled by several groups of investigators
in the past.  Subtraction of a template galaxy spectrum that has no
comparable emission has proved to be an effective technique. For our
analysis we have taken advantage of having in hand a significant number
of absorption--line galaxies, all observed under the same instrumental
conditions, in order to build such a template.  The spectral
characteristics of the galaxies used as templates are remarkably
similar to those of the LLAGN candidates. On average, the
difference between the velocity dispersion of the absorption lines of
our template galaxies and those of the LLAGN candidates is approximately 2 $\rm\AA$,
which is smaller than our resolution ($\sim 6 \rm\AA$). Figure 2 shows
the red part of the spectra of the LLAGN candidates after template subtraction.
Note the high [\ion{N}{2}]$\lambda$6584/H$\alpha$ ratios typical of
AGNs.
  
The classification of the activity type of all the 54 emission-line
galaxies in our sample was determined after subtraction of a galaxy
template. The emission--lines were measured by fitting gaussian
profiles. Our classification is based on the characteristic
line ratios shown by galaxies of different activity classes (Baldwin et
al. 1981; Veilleux \& Osterbrock 1987).  The emission--line galaxies in our sample
were separated into 3 groups: starburst galaxies, AGNs and LLAGNs.

The first diagnostic diagram, presented in Figure 3, is the diagram of
[\ion{N}{2}]$\lambda$6584/H$\alpha$ vs.
[\ion{O}{3}]$\lambda$5007/H$\beta$ (Baldwin et al. 1981; Veilleux \&
Osterbrock 1987). The distinction between AGN and starburst galaxies is
based on the empirical separation proposed by Veilleux \& Osterbrock
(1987). The separation between LINER and
Seyfert 2 galaxies corresponds to log([\ion{O}{3}]$\lambda5007$/H$\beta)
\leq 0.4$, as proposed by Coziol (1996). The same criterium is used to
distinguish between the low--excitation Starburst Nucleus Galaxies
(SBNGs) and the high--excitation HII galaxies.
This diagnostic diagram confirms the AGN nature of all our
LLAGN candidates. It also indicates that the majority of these
galaxies are LINER galaxies.

To double check our classification we also examined the ratio
[\ion{O}{1}]$\lambda$6300/H$\alpha$.  As was shown by Baldwin et al.
(1981), the presence and the strength of the [\ion{O}{1}]$\lambda$6300
are important factors in distinguishing between starburst galaxies and
AGNs. This is because this line is produced only in regions of
partial ionization which are much more extended in AGNs than in normal
HII regions.  For this reason the [\ion{O}{1}] line is rarely observed
and is always weaker in starburst galaxies than in AGNs.  In our sample,
the [\ion{O}{1}] line  is detected in 61\% of the starburst galaxies,
88\% of the LLAGNs and 60\% of the AGNs.  It is
interesting to note that the detection rate of [\ion{O}{1}] in the
starburst galaxies is much higher than usually observed in other
samples of starburst nucleus galaxies (generally of the order of 40\%
or less; see Coziol et al. 1997).

In Figure 4, we present the diagnostic diagram of the ratio
[\ion{O}{1}]$\lambda$6300/H$\alpha$ vs.
[\ion{O}{3}]$\lambda$5007/H$\beta$. The vertical line is the
[\ion{O}{1}] strength limit as proposed by Baldwin et al. (1981) to
separate the starburst galaxies from the AGNs.  This diagram confirms
the nature of the LLAGNs as determined using the previous diagnostic
diagram.  Only one candidate (otherwise a seyfert galaxy) has an
[\ion{O}{1}] line with strength comparable to those of starburst
galaxies.  Three other galaxies have values with error bars that cross the
boundary between the two activity classes. The fact that many starburst
galaxies in our sample have an unusually strong [\ion{O}{1}] line is
consistent with previous observations suggesting that a small fraction
of these galaxies show the simultaneous characteristics of an AGN
and an HII region (V\'eron et al. 1996). Our result suggests that
very few AGNs present the same ambiguity.
 
In Table 2, we list the results of our classification.  The name of the
objects are the same as in Table 1.  Columns 2, 3 and 4 give the ratios
[\ion{O}{3}]$\lambda$5007/H$\beta$, [\ion{N}{2}]$\lambda$6584/H$\alpha$
and [\ion{O}{1}]$\lambda$6300/H$\beta$ respectively.  The uncertainties
in these ratios were determined using Poisson statistics.  
In column 5, the different types of activity 
are described as: starburst galaxies (SBNGs or HII), AGNs 
(Seyfert 2: Sy2 or LINER: LNR) and LLAGNs (dSy2 or dLNR). Column 7 give the 
FWHM of the H$\alpha$ line. It may be interesting to note
that no Seyfert 1 galaxies was found in the groups.

\subsection{The low--luminosity nature of the hidden AGNs}

Rubin et al. (1991) already
noted that all the elliptical and the S0 galaxies in the HCGs have
ionized gas. They also remarked that this phenomenon is commonly
observed in samples of early--type galaxies (PJDSB).
But because the emission is hidden in the strong stellar
continuum the nature of the activity in these galaxies was not clearly
established. The subtraction of a template galaxy allow us 
to construct two different diagnostic diagrams which confirm the AGN nature of
these objects.  We will now verify that the AGNs in these galaxies
also have a low luminosity.  This will be done by comparing our candidates
with the low-luminosity AGNs discovered by PJDSB.

Because we do not have the calibrated spectra of our candidates we
cannot compare their fluxes directly with those of luminous AGNs.  To
test our assumption we use instead the EW of the
[\ion{N}{2}]$\lambda$6584 line, which, by definition, is proportional
to the ratio of the flux in the line divided by the stellar continuum
near the line.  In Figure 5 we compare the EW of the
[\ion{N}{2}]$\lambda$6584 line as a function of the ratio
[\ion{N}{2}]$\lambda$6584/H$\alpha$ for all the galaxies in our sample
with those of the low--luminosity AGNs observed by PJDSB. A very clear
pattern is seen:  the LLAGNs have higher
[\ion{N}{2}]$\lambda$6584/H$\alpha$ ratios and lower
EW([\ion{N}{2}]$\lambda$6584) than the luminous AGNs and the starburst
galaxies. Most important, the values observed for the LLAGNs are
identical to those of the low--luminosity AGNs as studied by PJDSB.  It is
important to realize that in this diagram there is only one way to
obtain the lowest EW value possible: the flux in the line must be low
and the stellar continuum must be high. In early-type galaxies, as in
the galaxies of the sample of PJDSB and also in our candidates (cf.
figure 7), the stellar continuum is already near maximum, which means
therefore that the line flux in the galaxies of these two samples must
be similar.  PJDSB determined that the characteristic H$\alpha$
luminosity of the low--luminosity AGN  in their sample is of the
order of 10$^{39}$ ergs s$^{-1}$ or lower. Then, by comparison, the
luminosity of the AGNs in our candidates must be of the same order,
which therefore confirms the low--luminosity nature of the AGNs in our
objects.  

\section {Discussion}

In Table 3 we present the fraction of each type of activity observed in
our sample of 82 galaxies. 
The LLAGNs form more than 50\% of the total AGN
population found in the whole sample. 
From the 82 galaxies studied here, only 67
galaxies are real group members (de Carvalho et al. 1997).  In Table 3
it can be seen that when only group members are considered, the
fraction of starburst galaxies drops and the fraction for the other
types slightly increases.
 
Going one step further, we can distinguish between the galaxies that
reside in the ``core'' and the ``halo'' of the groups. As discussed by
Ribeiro et al. (1997) most of HCG's are embedded in larger structures
forming extended haloes around more dense and dinamically distinct
cores. As shown by Ribeiro et al. these cores may be usefully defined
as the circular region around the group baricenter having mean surface
brightness $\rm \mu_B = 27\ mag\ arcsec^{-2}$, which is almost the same
limit used by Hickson (1982) as part of his criteria for constructing
his catalogue of compact groups. Table 4 shows how the 67 galaxies in
the groups distribute between the core (53 galaxies) and the halo (14
galaxies).  It is important to note that nearly all the AGNs, that is
the LLAGNs as well as the luminous AGNs, reside in the core of the
groups.

The fraction of each activity type encountered in the core regions is
reported in Table 3.  It can be seen that the fraction of starburst
galaxies decreases significantly in the core of the groups, where
AGNs (47.2\%) and non-emission line galaxies (37.7\%) dominate.
Because the star formation rate in all of these galaxies is
relatively low, we further conclude that the star formation rate in the
groups is generally at a low level.
      
Our sample is obviously biased towards galaxies which have high S/N
ratios, therefore favoring the most luminous galaxies in the groups. To
verify how this bias affects our conclusions on the spatial segregation
of the different activity types in the groups, we compare in Figure 6
the absolute magnitude of all the galaxies in our sample.  In this
figure we can see that the luminosity distributions for the LLAGNs
and the luminous AGNs are clearly biased towards higher values.  
However Figure 6 also shows that both the non emission--line galaxies
and the starburst galaxies show a comparable distribution of
luminosities.  Since these two types of galaxies have different spatial
locations in the groups, we may conclude that the segregation of
activity types in the groups is not due to a luminosity bias but
corresponds to a real physical effect. The fact that AGNs reside in
the most luminous galaxies of the groups is consistent with studies of
AGNs in the field which show that the probability of finding an AGN
increases with the luminosity of the host galaxy (PJDSB; Veilleux et
al. 1995).

In Figure 7 we present the distribution of morphological types of all the
galaxies in our sample. The LLAGNs reside either in elliptical or early--type
spirals whereas the luminous AGNs seem to reside mostly in early--type
spirals.  Figure 7 also shows that all the non-emission galaxies are
early-type galaxies (E, S0 or Sa). Consequently, the LLAGNs represent
36\% (41\% in the core) of the early--type population in the groups.
This result is consistent with the fraction of LLAGNs discovered by
PJDSB in their sample of luminous E and S0 galaxies.

\section {Conclusion}

We have shown that a significant fraction of the brightest galaxies of
compact groups display some sort of activity, either starburst or AGN.
Moreover, we find that about half of the AGN population is made of
LLAGNs, implying that AGNs are the most frequent activity 
types encountered in the groups.

We have further found that the AGNs -- both low--luminosity and normal --
reside in the most luminous galaxies of the groups and the
most early-type ones. These two characteristics are not unique to
compact groups but are consistent with the fact that the probability of
finding an AGN increases with the luminosity of the galaxies and that
low--luminosity AGNs are very common in early-type galaxies.  However,
we have also found that the AGNs tend to concentrate in the cores of
the groups. This may be viewed as a consequence of the luminosity and/or a
morphological segregation already found in the groups (Ribeiro et al.
1997). In other words, we have discovered a high fraction of AGNs in
the cores of the groups because the most luminous and early-type
galaxies are more concentrated in the groups and because the
probability of finding an AGN is higher in luminous and early-type
galaxies.

That the group produces some physical effects is even more obvious when
we consider the starburst galaxies:  the fraction of starburst galaxies
drops drastically when we go from the halo to the core of the groups. It
would be interesting to see if this behavior for the starburst galaxies
is due to a morphological bias. In Figure 7 we can see that
the starburst galaxies of our sample prefer late--type spirals.
Although suggestive, this result should be viewed with caution given
the scarcity of the data for these galaxies (less than 20\% of the
morphologies determined in our sample).
  
As it is, our discovery suggests that we do have, at least for the
core, a sort of density--morphology--activity relation. This relation
may be reminiscent of the density--morphology relation observed in
denser agglomerations of galaxies. For the HCGs, this relation would
suggest that the galaxies in the groups have a common history and that
their evolution was influenced by their environment.
 
As a working hypothesis we propose the following scenario for galaxy
evolution in the HCGs. Massive galaxies form by subsequent mergers of
smaller mass systems made of gas and stars.  Obviously, denser
environments will accelerate this process and the more massive galaxies 
will form first in the richest regions. The frequency and intensity of
mergers determine the morphologies, the most massive galaxies evolving
towards the most early--type morphologies. This would explain why in
HCGs the most luminous galaxies are also the most concentrated and the most early--type.
 
This formation scenario can incorporate a sequence where 
the activity level changes with time. The first phase is characterized
by a starburst and a Seyfert. Each merger triggers a starburst. 
When the mass of the galaxy is
sufficiently high or when a large amount of gas is available, the gas
falls into the center of the galaxy and form or nourish an AGN.  As the
starburst fades, depending on the reservoir of gas available and on the
rate of accretion, it remains either a Seyfert or a LINER.  Finally,
when the matter available to feed the AGN decreases, it remains a
LLAGN or a normal galaxy. 
This scenario would explain,
for instance, the difference observed in the groups between the
morphologies of the LLAGNs and the luminous AGNs. 
It would also explain why the core of the groups 
are dominated by a high fraction of LLAGNs and non--active early--type galaxies, 
since denser environments accelerate the process.  
The galaxies in the cores of
the groups are more evolved than in the outer regions. 
This would explain why starburst galaxies are observed in higher number
in the halo than in the core. This scenario may also explain why we do 
not find any Seyfert 1 galaxy in the core, as they would 
be visible only at the beginning of the process. 

Following this scenario, most of the activity
in HCGs would have taken place sometime in the recent past, and the
groups are now observed at a quiet phase of their activity.

\acknowledgments

We thank Natalie Stout for his careful reading of the manuscript. 
R. Coziol acknowledges the financial
support of FAPESP under contract 94/3005-0 (while working 
at the Instit\'uto Nacional de Pesquisas Espacias/INPE) and of CNPq, 
under contracts 360715/96-6 (NV).
For his part, A.L.B. Ribeiro acknowledges the support of the Brazilian CAPES.

\clearpage
\begin{deluxetable}{cccccl}
\footnotesize
\tablecaption{Characteristics of the low--luminosity AGNs \label{tbl-1}}
\tablewidth{0pt}
\tablehead{
\colhead{HCG \#}&\colhead{Hickson} &
\colhead{V$_{\rm o}$} & \colhead{B}   & \colhead{M$_{\rm B}$}  & \colhead{Hubble}\\
\colhead{}      &\colhead{id.}     &
\colhead{km s$^{-1}$} & \colhead{}   & \colhead{}  & \colhead{Type} 
}
\startdata
22--1 &   a     & 2681 & 12.6 &  -20.2 &  E2     \nl
22--5 &   d     & 9268 & 15.2 &  -20.3 &  E3     \nl
22--6 &   e     & 9611 & 15.2 &  -20.3 &  E5     \nl
23--3 &   a     & 4869 & 15.2 &  -18.8 &  Sab    \nl
23--5 &   c     & 5373 & 16.1 &  -18.1 &  S0     \nl
40--1 &   a     & 6634 & 14.3 &  -20.4 &  E3     \nl
42--1 &   a     & 3712 & 12.6 &  -20.8 &  E3     \nl
62--1 &   a     & 4259 & 13.3 &  -20.5 &  E3     \nl
64--1 & \nodata & 6264 & 13.8 &  -20.8 & \nodata \nl
67--11& \nodata & 6614 & 16.5 &  -18.2 & \nodata \nl
86--1 &   a     & 6013 & 14.2 &  -20.3 &  E2     \nl
86--3 &   b     & 5863 & 14.8 &  -19.7 &  E2     \nl
87--1 &   a     & 8436 & 14.8 &  -20.4 &  Sbc    \nl
87--3 &   b     & 8738 & 15.4 &  -19.9 & S0      \nl
88--1 &   a     & 6007 & 13.7 &  -20.8 & Sb      \nl
88--2 &   b     & 6124 & 13.8 &  -20.8 & SBb     \nl
\enddata 
\end{deluxetable}

\clearpage
\begin{deluxetable}{lccclc}
\footnotesize
\tablecaption{Spectral characteristics of HCG's galaxies\label{tbl-2}}
\tablewidth{0pt}
\tablehead{
\colhead{HCG  \#} & \colhead{$\frac{{\rm [OIII]}\lambda50007}{{\rm H}\beta}$} &
\colhead{$\frac{{\rm [NII]}\lambda6548}{{\rm H}\alpha}$} & 
\colhead{$\frac{{\rm [OI]}\lambda6300}{{\rm H}\alpha}$}  & \colhead{Activity} &\colhead{FWHM} \\
\colhead{}        &\colhead{}                                                 &
\colhead{}                                               &
\colhead{}                                               &\colhead{Type}     &\colhead{}       \\ 
\colhead{}        &\colhead{}                            &
\colhead{}        &\colhead{}                            &\colhead{}          &\colhead{(km s$^{-1}$)} 
}
\startdata
04  01 & -0.26            &  -0.31             &  -1.93             &  SBNG            &  535  \nl
04  03 & -0.37            &  -0.32             &  -1.61             &  SBNG            &  367  \nl
04  11 &  0.77            &  -1.26             &  -1.96             &  HII             &  327  \nl
16  01 &  0.31 $\pm$ 0.02 &   0.09             &  -0.80 $\pm$ 0.04  &  LNR             &  498  \nl
16  02 &  0.58 $\pm$ 0.01 &   0.05             &  -0.40 $\pm$ 0.01  &  Sy2             &  816  \nl
16  03 &  0.12 $\pm$ 0.01 &  -0.35             &  -1.56 $\pm$ 0.18  &  SBNG            &  499  \nl
16  04 & -0.35            &  -0.38             &  -1.91 $\pm$ 0.13  &  SBNG            &  489  \nl
16  05 &  0.35 $\pm$ 0.01 &  -0.22             &  -1.14 $\pm$ 0.04  &  LNR             &  533  \nl
16  06 &  0.61 $\pm$ 0.03 &  -0.8  $\pm$ 0.1   &  -1.07 $\pm$ 0.31  &  HII             &  383  \nl
19  05 & -0.12            &  -0.495            &   \nodata          &  SBNG            &  424  \nl
22  01 &  0.50 $\pm$ 0.03 &   0.03 $\pm$ 0.02  &  -0.9  $\pm$ 0.3   &  dSy2            &  953  \nl
22  05 &  0.50 $\pm$ 0.08 &   0.42 $\pm$ 0.02  &  -0.7  $\pm$ 0.4   &  dSy2            &  437  \nl
22  06 &  0.64 $\pm$ 0.08 &   0.23 $\pm$ 0.02  &  -0.7  $\pm$ 0.2   &  dSy2            &  518  \nl
22  07 & -0.5  $\pm$ 0.1  &  -0.34 $\pm$ 0.02  &   \nodata          &  SBNG            &  $<$360\nl
22  08 & \nodata          &  -0.11 $\pm$ 0.01  &   \nodata          &  LNR?            &  533  \nl
23  02 &  0.79            &   0.74             &  -0.11             &  Sy2             &  716  \nl
23  03 &  0.30            &   0.21             &  -0.49             &  dLNR            &  542  \nl
23  04 & -0.25            &  -0.39             &  -1.78             &  SBNG            &  480  \nl
23  05 &  0.47            &  -0.09             &  -2.0  $\pm$ 0.1   &  dSy2            &  796  \nl
23  06 & -0.21            &  -0.50             &   \nodata          &  SBNG            &  356  \nl
40  01 &  0.0  $\pm$ 0.1  &  -0.06 $\pm$ 0.03  &   \nodata          &  dLNR            &  560  \nl
40  04 &  0.1  $\pm$ 0.2  &  -0.19 $\pm$ 0.03  &  -0.9  $\pm$ 0.3   &  LNR             &  730  \nl
40  05 &  0.43 $\pm$ 0.05 &  -0.22 $\pm$ 0.02  &   \nodata          &  Sy2             &  776  \nl
42  01 & \nodata          &   0.25 $\pm$ 0.01  &  -1.0  $\pm$ 0.5   &  dLNR            &  765  \nl
48  19 & -0.29 $\pm$ 0.03 &  -0.54 $\pm$ 0.01  &  -1.4  $\pm$ 0.3   &  SBNG            &  392  \nl
48  25 &  0.36 $\pm$ 0.01 &  -0.96 $\pm$ 0.07  &   \nodata          &  SBNG            &  366  \nl
62  01 &  0.00 $\pm$ 0.05 &  -0.22 $\pm$ 0.01  &  -1.1  $\pm$ 0.2   &  dLNR            &  795  \nl
63  04 &  0.11 $\pm$ 0.04 &  -0.18 $\pm$ 0.01  &   \nodata          &  LNR             &  588  \nl
63  06 & -0.49 $\pm$ 0.06 &  -0.36 $\pm$ 0.01  &   \nodata          &  SBNG            &  512  \nl
64  01 &  0.07 $\pm$ 0.05 &   0.42 $\pm$ 0.03  &  -0.8  $\pm$ 0.7   &  dLNR            &  701  \nl
64  22 & -0.23 $\pm$ 0.08 &  -0.48 $\pm$ 0.02  &  -1.1  $\pm$ 0.2   &  SBNG            &  368  \nl
67  02 & -0.15 $\pm$ 0.07 &  -0.47 $\pm$ 0.03  &   \nodata          &  SBNG            &  658  \nl
67  06 & -0.16 $\pm$ 0.04 &  -0.43 $\pm$ 0.03  &   \nodata          &  SBNG            &  354  \nl
67  11 &  0.07 $\pm$ 0.14 &   0.71 $\pm$ 0.09  &   0.3  $\pm$ 0.2   &  dLNR            &  677  \nl
86  01 & -0.40 $\pm$ 0.19 &  -0.21 $\pm$ 0.06  &  -0.6  $\pm$ 0.2   &  dLNR            &  402  \nl
86  03 &  0.95 $\pm$ 0.02 &  -0.08 $\pm$ 0.01  &  -1.1  $\pm$ 0.3   &  dSy2            &  947  \nl
86  04 &  0.17 $\pm$ 0.03 &  -0.15 $\pm$ 0.01  &  -0.83 $\pm$ 0.09  &  LNR             &  977  \nl
86  07 &  1.16 $\pm$ 0.01 &   0.09 $\pm$ 0.01  &  -0.37 $\pm$ 0.03  &  Sy2             &  728  \nl
86  08 & -0.35 $\pm$ 0.04 &  -0.47 $\pm$ 0.01  &  -1.5  $\pm$ 0.2   &  SBNG            &  352  \nl
86  09 & -0.10 $\pm$ 0.04 &  -0.66 $\pm$ 0.03  &   \nodata          &  SBNG            &  763  \nl
87  01 &  0.08            &  -0.08             &  -0.500$\pm$ 0.001 &  dLNR            &  601  \nl
87  02 & -0.34            &  -0.42             &  -1.75 $\pm$ 0.01  &  SBNG            &  522  \nl
87  03 & -0.35            &   0.11             &   \nodata          &  dLNR            &  508  \nl
87  04 & -0.38            &  -0.36             &  -0.76             &  SBNG            &  $<$360\nl
87  05 &  0.31            &  -0.60             &  -1.14             &  SBNG            &  358  \nl
87  07 & -0.08            &  -0.42             &  -1.65             &  SBNG            &  499  \nl
88  01 &  0.48            &  -0.05             &  -1.00             &  dSy2            &  704  \nl
88  02 &  0.18            &   0.28             &  -1.11             &  dLNR            &  517  \nl
88  07 &  0.10            &  -0.49             &   \nodata          &  SBNG            &  231  \nl
90  01 &  0.97$\pm$ 0.02  &   0.08 $\pm$ 0.01  &  -1.3  $\pm$ 0.8   &  Sy2             &  470  \nl
90  04 &  0.02$\pm$ 0.07  &  -0.20 $\pm$ 0.01  &  -1.1  $\pm$ 0.2   &  LNR             &  441  \nl
90  09 &  0.21$\pm$ 0.01  &  -0.61 $\pm$ 0.03  &  -1.4  $\pm$ 0.5   &  SBNG            &  359  \nl
97  05 &  0.23$\pm$ 0.06  &  -0.28 $\pm$ 0.04  &   \nodata          &  LNR             &  397  \nl
97  06 &  0.40$\pm$ 0.04  &  -0.26 $\pm$ 0.03  &   \nodata          &  Sy2             &  394  \nl
\enddata 
\end{deluxetable}

\clearpage
\begin{deluxetable}{cccccccl}
\footnotesize
\tablecaption{Fraction of activity types \label{tbl-3}}
\tablewidth{0pt}
\tablehead{
\colhead{}   & \colhead{all}  & \colhead{in group} & \colhead{in core}    
}
\startdata
No of Galaxies     & 82   & 67     & 53    \nl
No emission        & 34\% & 37.3\% & 37.8\%  \nl
LLAGNs             & 20\% & 20.9\% & 24.5\%  \nl
AGNs               & 18\% & 19.5\% & 22.6\%  \nl
Starbursts         & 28\% & 22.3\% & 15.1\%  \nl
\enddata 
\end{deluxetable}

\clearpage
\begin{deluxetable}{cccccccccc}
\footnotesize
\tablecaption{Classification of activity type \label{tbl-4}}
\tablewidth{0pt}
\tablehead{
\colhead{HCG} & \multicolumn{4}{c}{core} &\colhead{}
& \multicolumn{4}{c}{halo} \\
\cline{2-5} \cline{7-10} \\
\colhead{\#}  & \colhead{(1)}  & \colhead{(2)} & \colhead{(3)}  
& \colhead{(4)} &\colhead{} & \colhead{(1)}  & \colhead{(2)} & \colhead{(3)}  & \colhead{(4)}
}
\startdata
04&  \nodata    & \nodata    & 1a      & \nodata     & & \nodata & \nodata & \nodata    & \nodata \nl
16&  1a, 2b, 5d & \nodata    & 3, 4c   & \nodata     & & \nodata & \nodata &  6         & \nodata \nl
19&  \nodata    & \nodata    & \nodata & 1a          & & \nodata & \nodata & \nodata    & \nodata \nl
22&  \nodata    & 1a         & \nodata & 3, 4b       & & \nodata & \nodata & \nodata    & \nodata \nl
23&  \nodata    & 3a, 5c     & 4d      & \nodata     & & \nodata & \nodata & \nodata    & \nodata \nl
40&  4d,5e      & 1a         & \nodata & 2b          & & \nodata & \nodata & \nodata    & \nodata \nl
42&  \nodata    & 1a         & \nodata & 2b, 4c      & & \nodata & \nodata & \nodata    & \nodata \nl 
48&  \nodata    & \nodata    & \nodata & 1a          & & \nodata & \nodata &  25        & \nodata \nl
62&  \nodata    & 1a         & \nodata & 2b, 3c, 4   & & \nodata & \nodata & \nodata    & 8       \nl
63&  4a         & \nodata    & \nodata & \nodata     & & \nodata & \nodata &   6        & \nodata \nl
64&  \nodata    & 1          & \nodata & \nodata     & & \nodata & \nodata &  22        & \nodata \nl
67&  \nodata    & \nodata    & 2b      & 1a, 3       & & \nodata & 11      &   6        & \nodata \nl
86&  4c         & 1a, 3b     & \nodata & 6d          & & 7       & \nodata &   8, 9     & 2,10    \nl
87&  4c         & 1a, 3b     & 2, 5    & \nodata     & & \nodata & \nodata & \nodata    & \nodata \nl
88&  \nodata    & 1a, 2b     & 7       & \nodata     & & \nodata & \nodata & \nodata    & \nodata \nl
90&  1a, 4d     & \nodata    & 9       & 2b, 3c, 6   & & \nodata & \nodata & \nodata    & \nodata \nl
97&  5, 6b      & \nodata    & \nodata & 2, 4c, 3d   & & \nodata & \nodata & \nodata    & 10, 11  \nl
\enddata 
\tablenotetext{}{Activity types: (1) = AGN; (2) = LLAGN; 
\phm{a}(3) = HII-SBNG; (4) = no emission.}
\end{deluxetable}

\clearpage

\figcaption[f1a.eps,f1b.eps]{Optical spectra of the LLAGN candidates.
The spectra are in the rest frame of the object.
The flux scale was normalized to one and the spectra are shifted for presentation.
Note the similarity of all the spectra. 
Nuclear activity is suggested by the presence of the Nitrogen emission line 
([NII]$\lambda$6584). }

\figurenum{2}
\figcaption[f2a.eps,f2b.eps]{ The spectra of the LLAGN candidates
after subtraction of a template galaxy.
The galaxies are shown in the same order as in Fig. 1.
The ratios [NII]$\lambda$6584/H$\alpha$ are high and 
[OI]$\lambda$6300 is detected in almost all the galaxies.}

\figurenum{3}
\figcaption[f3.eps]{Diagnostic diagram of emission--line ratio and classification
of the different activity of our sample of emission--line galaxies  
The dot-dashed line at log([OIII]/H$\beta$) = 0.4 establishes
a distinction between Seyfert 2 and LINERs (Coziol 1996). 
The solid line indicates 
the empirical separation between starburst galaxies
and AGNs as determined by Veilleux and Osterbrock (1987).}

\figurenum{4}
\figcaption[f4.eps]{Second diagnostic diagram of emission--line ratios.
The vertical dashed line at log([OI]/H$\alpha$) = 1.3 establishes
a distinction between Starbursts and AGNs (Baldwin et al. 1981).}

\figurenum{5}
\figcaption[f5.eps]{ Equivalent widths (EW) of the 
[NII]$\lambda$6584 line vs. its strength in comparison
to H$\alpha$. The LLAGN possess a high ratio of [NII]$\lambda$6584/H$\alpha$
and a low EW. The values for the LLAGN candidates in our sample are the same
as those of the LLAGN studied by Phillips et al. (1986).}

\figurenum{6}
\figcaption[f6.eps]{Distribution of the luminosities of the galaxies 
with different activity types in our sample.
The AGNs are located in the most luminous 
galaxies of the groups, but the starbursts and the non emission galaxies show 
comparable luminosities. The low fraction of starburst observed in the core 
cannot be explain by a luminosity bias.} 

\figurenum{7}
\figcaption[f7.eps]{Distribution of the morphologies of the galaxies 
with different activity type in our sample. The LLAGNs and 
the non emission line galaxies are located in the
most early--type galaxies of the groups. 
The luminous AGNs are mostly early--type spirals.}

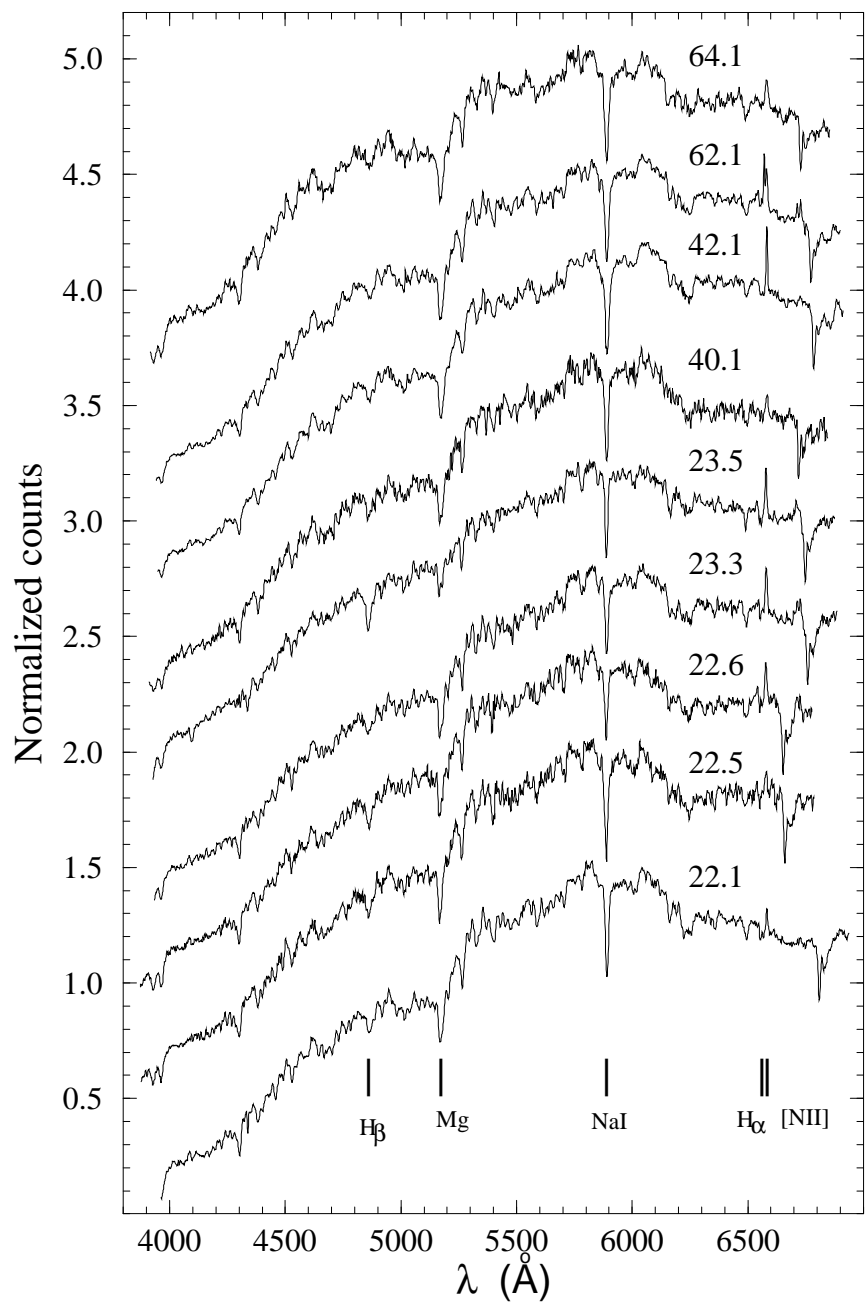
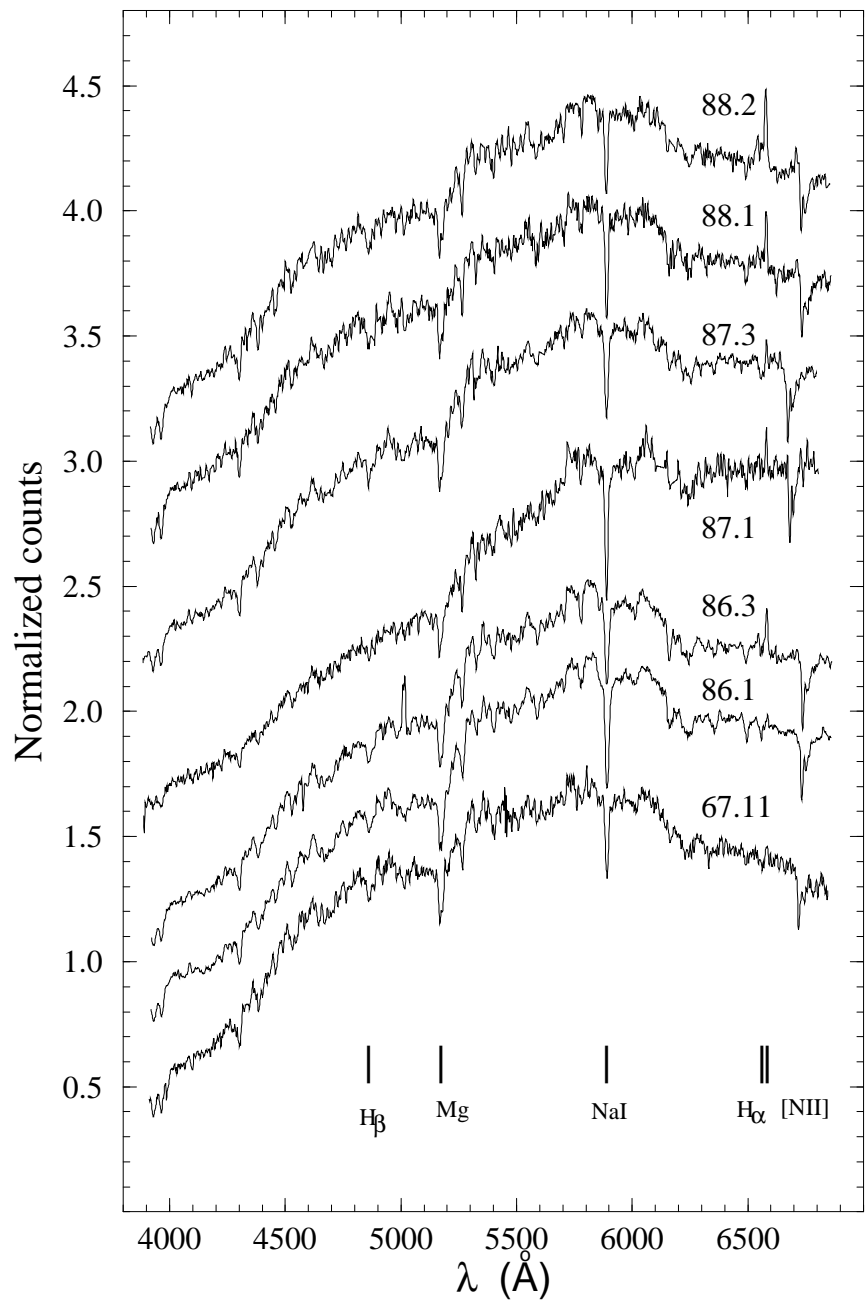
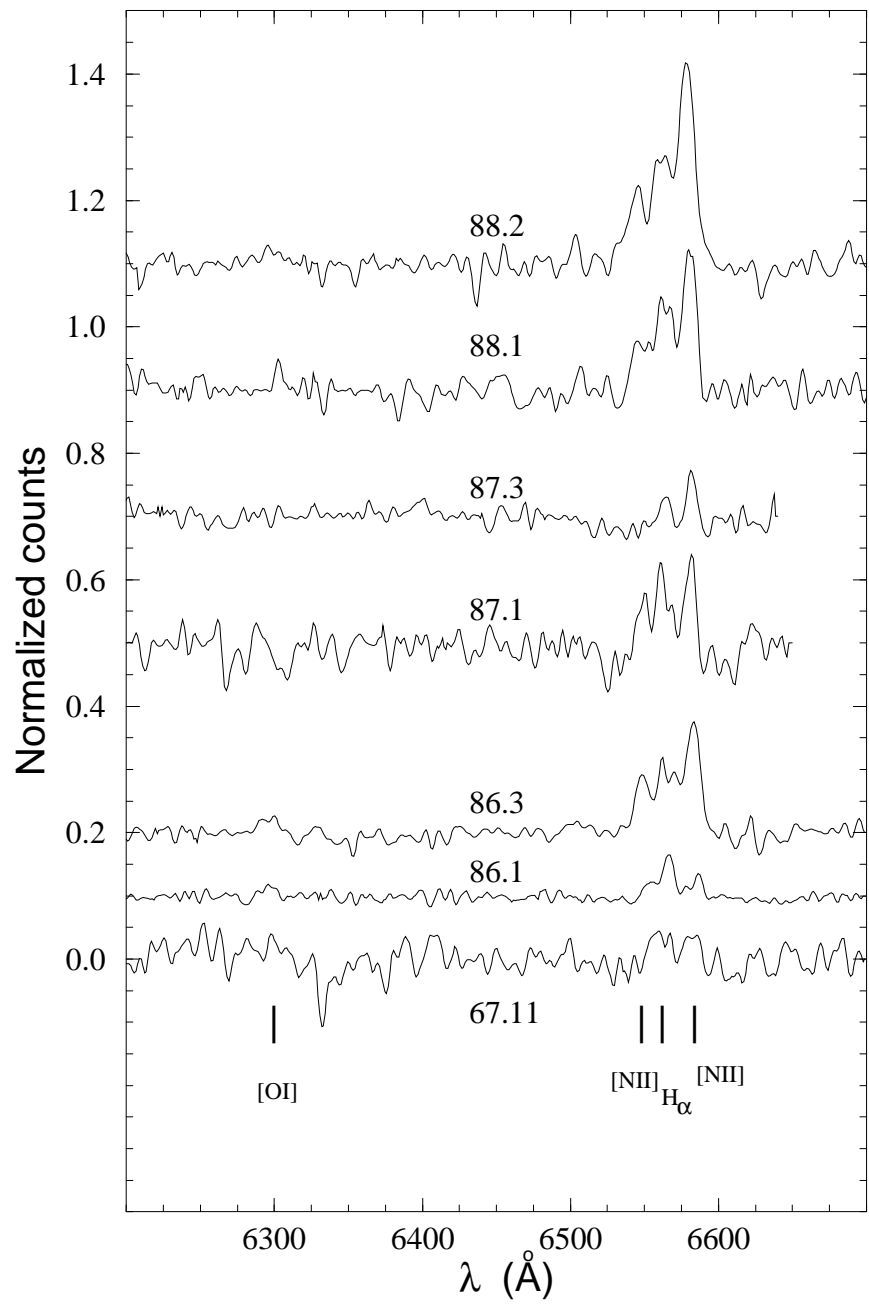
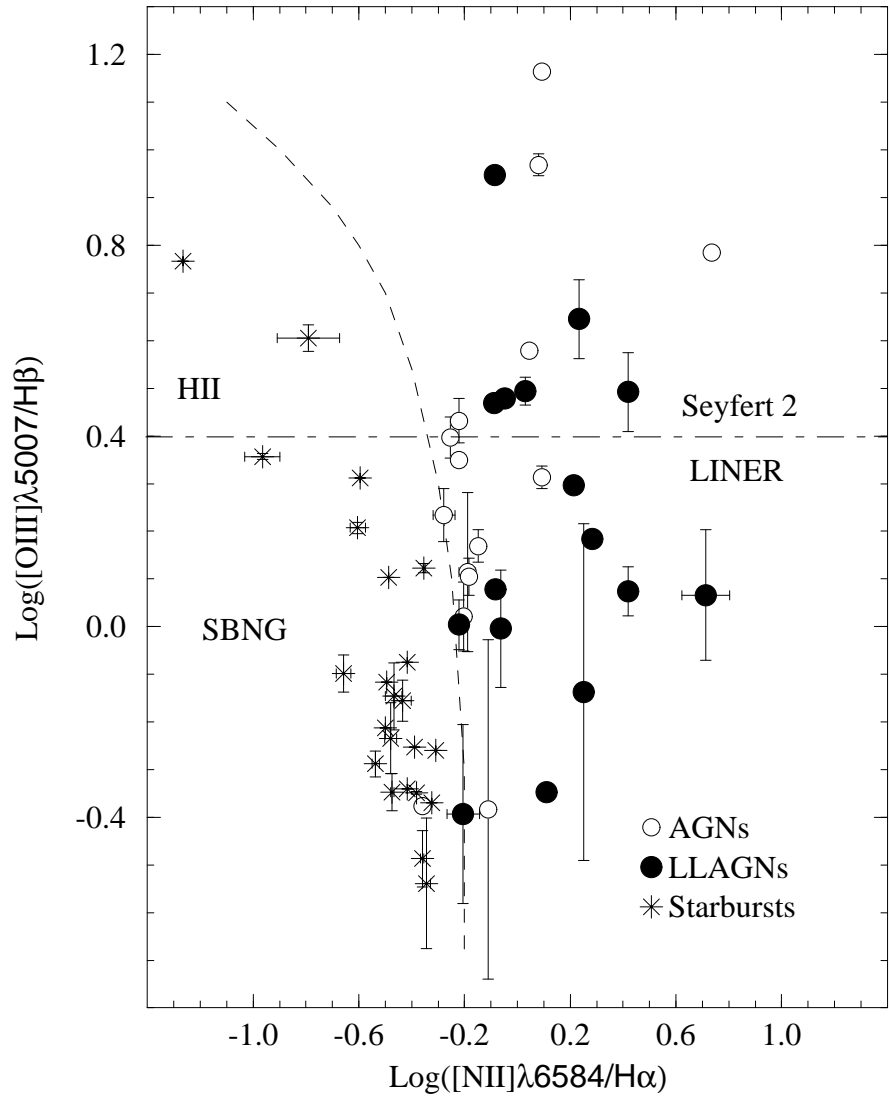
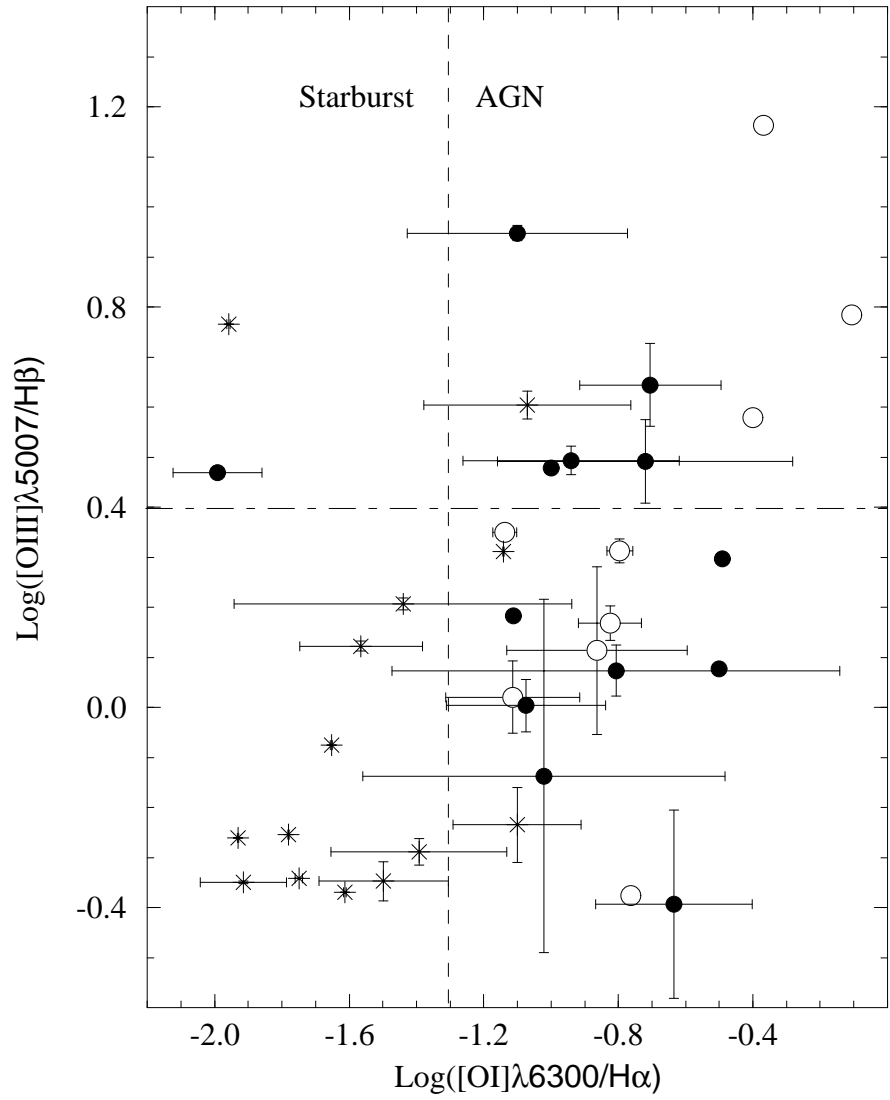
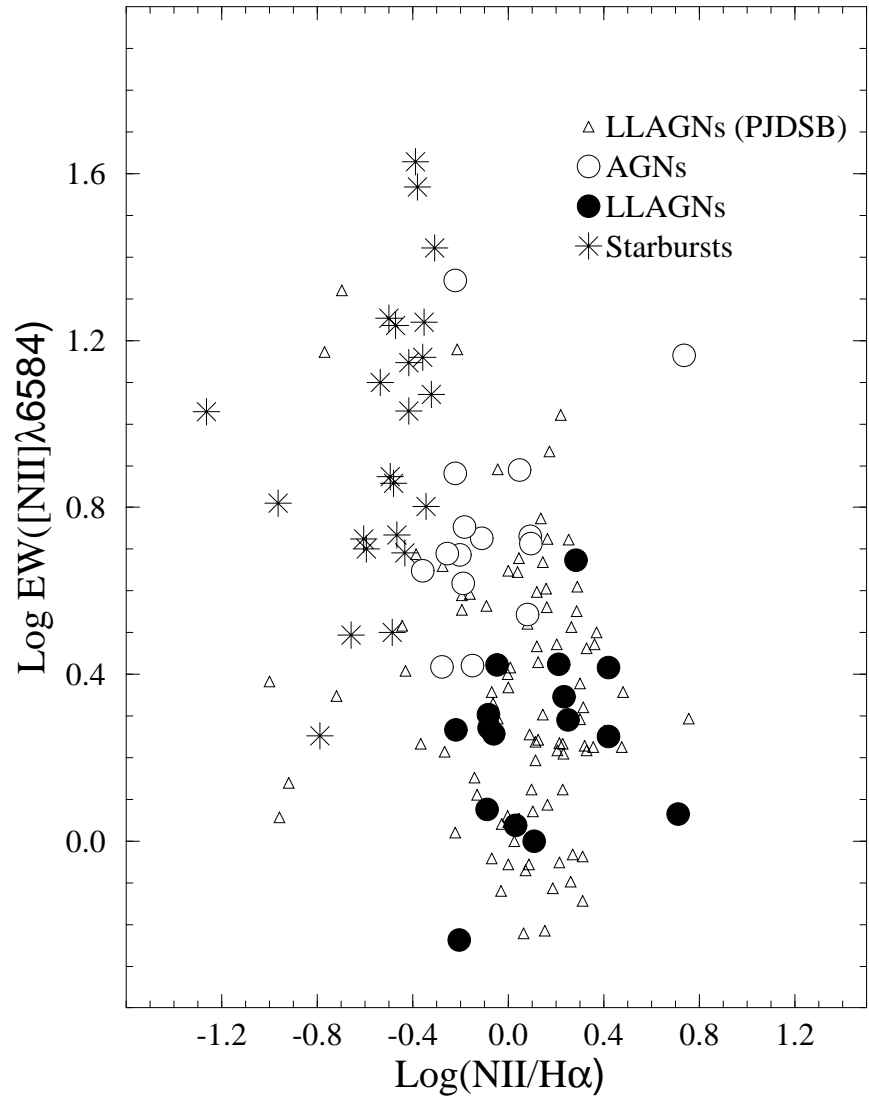
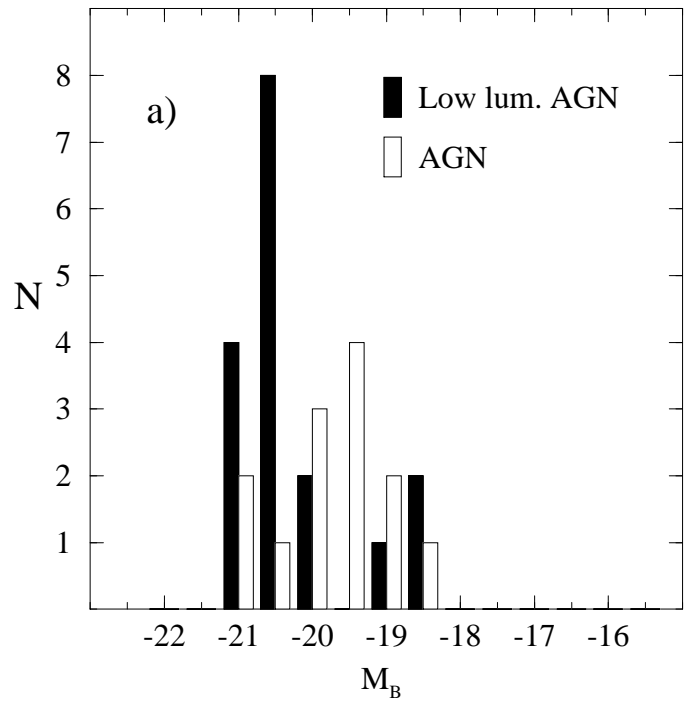
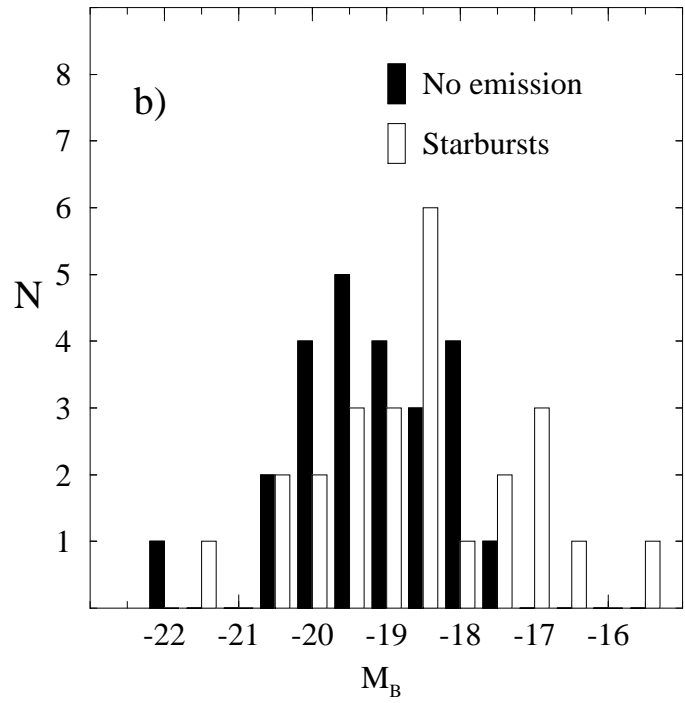
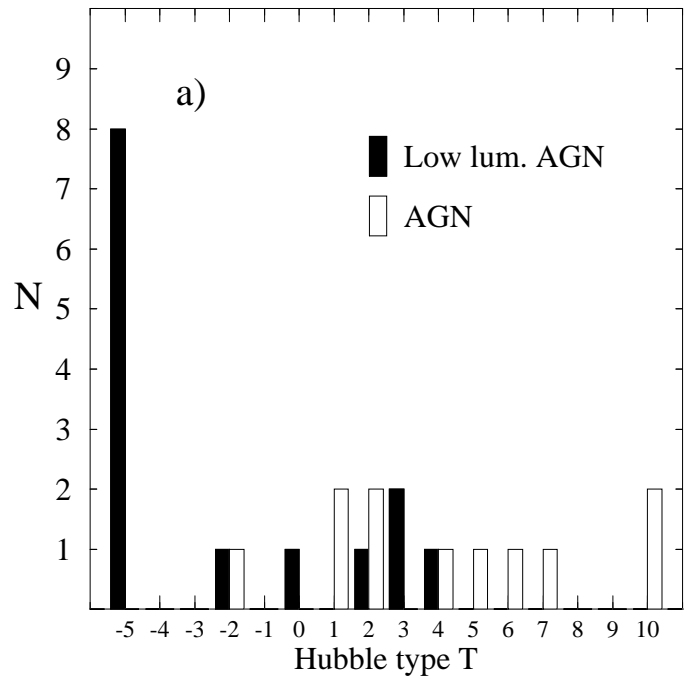
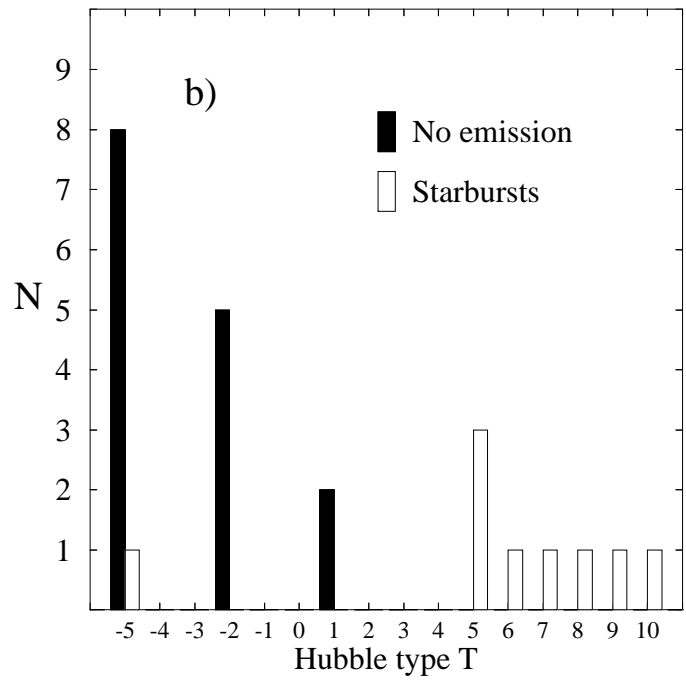
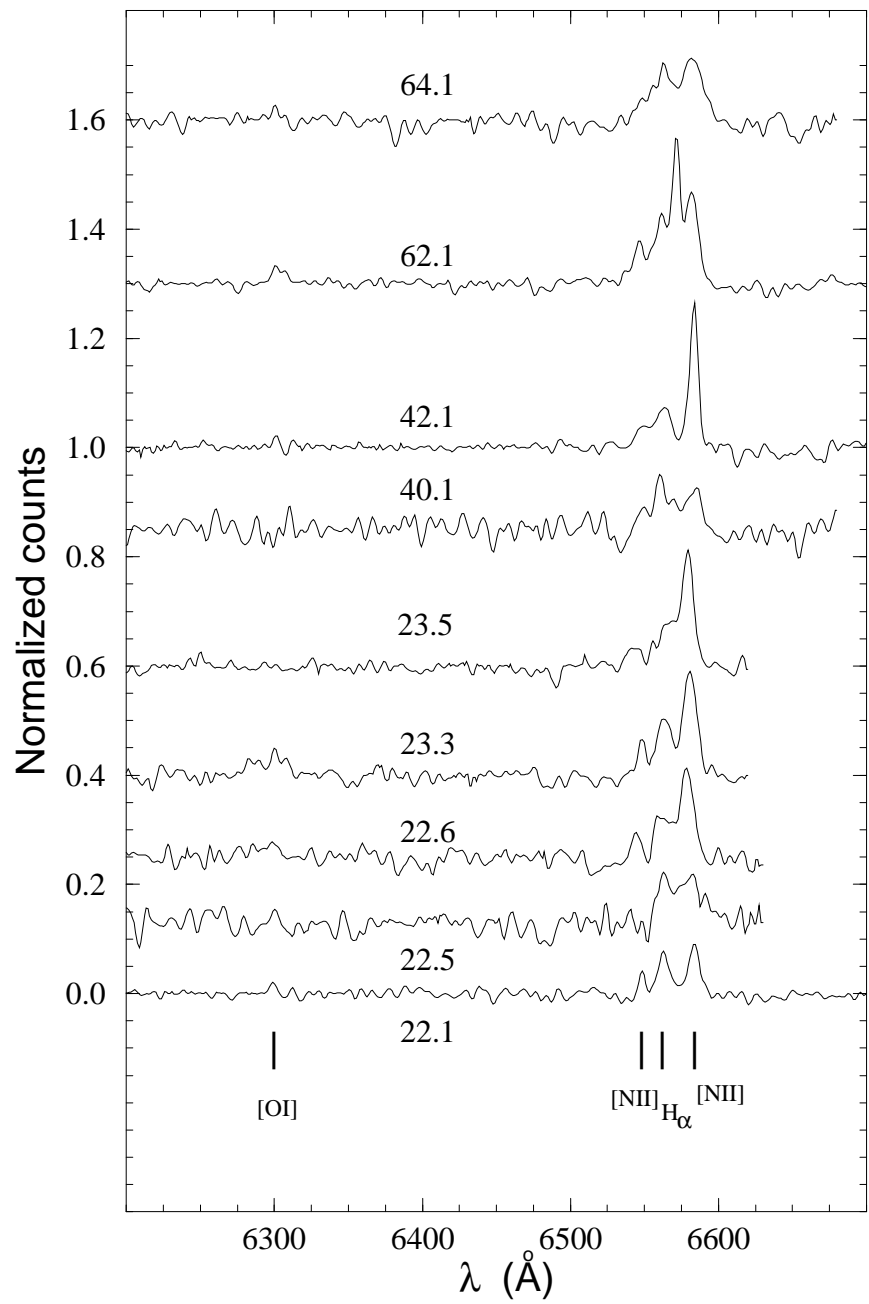
\end{document}